\documentclass[12pt,preprint]{emulateapj}

\shorttitle{GRB 100316D/SN 2010bh Host Metallicity}
\shortauthors{Levesque et al.}

\begin{document}

\title{Metallicity in the GRB 100316D/SN 2010bh Host Complex\footnotemark[1]} \footnotetext[1]{This paper includes data gathered with the 6.5 meter Magellan Telescopes
located at Las Campanas Observatory, Chile.}
\author{Emily M. Levesque$^{2,3}$, Edo Berger$^4$, Alicia M. Soderberg$^4$, Ryan Chornock$^4$}

\begin{abstract}
The recent long-duration GRB 100316D, associated with supernova SN 2010bh and detected by {\it Swift}, is one of the nearest GRB-SNe ever observed ($z = 0.059$).   This provides us with a unique opportunity to study the explosion environment on $\sim{\rm kpc}$ scale in relation to the host galaxy complex.  Here we present spatially-resolved spectrophotometry of the host galaxy, focusing on both the explosion site and the brightest star-forming regions.  Using these data, we extract the spatial profiles of the relevant emission features (H$\alpha$, H$\beta$, [OIII]$\lambda 5007$, and [NII]$\lambda 6584$), and use these profiles to examine variations in metallicity and star formation rate as a function of position in the host galaxy.  We conclude that GRB 100316D/SN2010bh occurred in a low-metallicity host galaxy, and that the GRB-SN explosion site corresponds to the region with the lowest metallicity and highest star formation rate sampled by our observations.
\end{abstract}

\section{Introduction}
\footnotetext[2]{CASA, Department of Astrophysical and Planetary Sciences, University of Colorado 389-UCB, Boulder, CO 80309, USA; \texttt{Emily.Levesque@colorado.edu}}
\footnotetext[3]{Einstein Fellow}
\footnotetext[4]{Harvard-Smithsonian Center for Astrophysics, 60 Garden St., Cambridge, MA 02138, USA}

Long-duration gamma-ray bursts (LGRBs) arise from the core-collapse deaths of some massive stars.  The primary line of evidence for this association comes from the detection of supernova signatures following the GRB.  These include photometric re-brightening with the magnitude, color, shape, and timescale expected for a supernova (e.g. Bloom et al.\ 1999, Zeh et al.\ 2004, Soderberg et al.\ 2006, Woosley \& Bloom 2006, Cobb et al.\ 2010), and more importantly from the spectroscopic detection of broad-lined Type Ic supernovae (SNe) in several LGRBs at $z\lesssim 0.3$ (e.g., Galama et al.\ 1998, Stanek et al.\ 2003, Malesani et al.\ 2004, Modjaz et al.\ 2006).  The connection with Type Ic SNe is extremely important, since these events are thought to be caused by the collapse of massive stars that have shed their outer hydrogen and helium envelopes (see Filippenko 1997).  Similarly, the broad absorption features indicate the presence of large ejecta velocities, and perhaps energies (e.g. Galama et al. 1998, Pian et al. 2006, Modjaz et al. 2008).

The most recent addition to this class of spectroscopically identified GRB-SNe is GRB 100316D/SN 2010bh at $z=0.059$ (Stamatikos et al.\ 2010, Vergani et al.\ 2010, Bufano et al.\ 2011).   In Chornock et al.\ (2011) we reported the spectroscopic discovery of the accompanying SN 2010bh, classified it as a broad-lined Type Ic SN based on optical and near-infrared spectra, and delineated the evolution of its photospheric velocity.  We also demonstrated that the host galaxy has a low luminosity, $L\approx 0.1$ L$_*$, and a low metallicity, $Z \lesssim 0.4Z_\odot$, similar to other GRB-SNe (e.g. Levesque et al.\ 2010a).  In addition, Starling et al.\ (2011) presented {\it Hubble Space Telescope} observations of the host galaxy, which revealed a complex morphology, possibly suggestive of recent merger activity.

The metallicities of the GRB 100316D/SN 2010bh host galaxy and explosion site are of particular interest.  In recent years, a connection has been suggested between LGRBs and low-metallicity host environments. LGRBs, associated with star-forming host environments, are observed in fainter and more irregular galaxies than their core-collapse SN counterparts (e.g. Fruchter et al.\ 2006, Wainwright et al.\ 2007). Their host galaxies, on average, fall below the luminosity-metallicity and mass-metallicity relations for star forming galaxies out to $z \sim 1$ (e.g. Modjaz et al.\ 2008, Kocevski et al.\ 2009, Levesque et al.\ 2010a,b).  However, the physical mechanism driving this apparent metallicity trend is still poorly understood.  LGRB host environments do not appear to adhere to any low-metallicity cut-off, with several high-metallicity host galaxies and explosion sites noted in the literature (Graham et al.\ 2009, Levesque et al.\ 2010b,c).  Similarly, the relativistic supernova SN 2009bb, found to have a comparable ejecta velocity to GRB-SNe but with no accompanying LGRB, was also observed in a high-metallicity host environment (Soderberg et al.\ 2010, Levesque et al.\ 2010d).  There is also no apparent correlation between host metallicity and gamma-ray energy release for LGRBs (Levesque et al.\ 2010e), a result that is at odds with previous predictions of the progenitor models (MacFadyen \& Woosley 1999).  The low-metallicity trend could also be an artifact of these eventsÕ young progenitor ages, the star formation histories of their hosts, or selection effects related to their star formation rates (e.g. Bloom et al.\ 2002, Berger et al.\ 2007, Kocevski \& West 2011).

One limitation of the aforementioned studies is their reliance on global metallicities.  For the majority of LGRB hosts at $z \gtrsim 0.3$, pinpointing the LGRB explosion site and acquiring site-specific spectra within the small, faint host galaxies is a difficult proposition. The nearby GRB-SNe therefore offer a unique opportunity to probe the metallicities of their host environments in detail.  For these spatially-resolved hosts we can determine metallicities at the LGRB host site as well as the surrounding star-forming regions of the galaxy, allowing us to pinpoint the precise environments that produce GRB-SNe and place these host sites in context with their global host galaxy properties. 

Here we present spatially-resolved spectroscopy of the GRB 100316D/SN 2010bh host galaxy.  These observations allow us to place the properties of the explosion site in the context of the broad host properties.  We discuss the observations and describe the analysis technique in \S 2.  We derive spatially-resolved line fluxes, line ratios, metallicities, and star-formation rates (SFRs) across the host galaxy, with a particular focus on the GRB-SN explosion site in \S 3. Finally, we consider the implications that these results for our understanding of LGRB host galaxies and the progenitor population in \S 4.  Throughout the paper we use the standard cosmological parameters $H_0=71$ km s$^{-1}$ Mpc$^{-1}$, $\Omega_m=0.27$, and $\Omega_\Lambda=0.73$.

\section{Observations and Reduction}
On 2010 May 8 UT we obtained a 2400 s spectrum of the GRB\,100316D/SN\,2010bh explosion site (hereafter, ``site''), using the Low Dispersion Survey Spectrograph (LDSS3) on the Magellan/Clay 6.5-m telescope at Las Campanas Observatory, using the 1" center slit; seeing was 0.88" (as measured from acquisition images in $r$) and the spectral resolution was 8.5\AA. A second spectrum was obtained on 2010 May 10 UT for a total exposure time of 1800 s, with the 1" blue slit aligned along the extended bright region of the host galaxy complex (hereafter, ``host''); seeing was 0.62" (as measured from acquisition images in $i$) and the spectral resolution was 9.0\AA. The slit positions for the ``site" and ``host" observations are shown in Figure 1 (top left and top right, respectively).  As the airmass was low ($\sim 1.4-1.5$) the impact on the spectra from not observing at the parallactic angle is expected to be minimal. We also acquired observations of the spectrophotometric standard stars EG 131 and LTT 3864 (Bessell et al.\ 1999, Hamuy et al.\ 1994) for flux calibration of the ``site" and ``host" spectra respectively.  All spectra were taken using the using the VPH-All grism, and cover the wavelength range 3800-7200\AA.

To locate the explosion site of GRB 100316D/SN 2010bh along the slit in the ``site'' spectrum we astrometrically aligned the $r$-band acquisition image with an $r$-band image of the GRB-SN from 2010 March 19.99 UT obtained with the same instrument.  Using 65 objects in common to the two images, we find an astrometric accuracy of $1\sigma=16$ mas in each coordinate.  This is significantly better than a single pixel ($0.19''$), but we expect an overall systematic uncertainty of about $\pm 1$ pixel due to the overall slit positioning.

We reduced the data using standard routines in IRAF\footnote{IRAF is distributed by NOAO, which is operated by AURA, Inc., under cooperative agreement with the NSF.}, including bias correction, flatfielding, and cosmic ray removal. To isolate the host galaxy emission lines, and to remove contamination from a foreground star in the bright southwestern region of the host complex in our ``site" observations (see Figure 1), we fit the continuum of both objects with a legendre polynomial using the IRAF task \texttt{background} and used this fit to subtract the continuum from our two-dimensional spectra. The results are shown in Figure 2.

We extracted line profiles along the spatial direction for four emission lines in each of our observed two-dimensional spectra: H$\beta$, [OIII]$\lambda 5007$, H$\alpha$, and [NII]$\lambda 6584$.  The bright H$\alpha$ and [OIII]$\lambda 5007$ features were traced using the IRAF task {\tt apall} with an optimal extraction algorithm that identified and rejected deviant pixels.  For the weaker [NII]$\lambda 6584$ and H$\beta$ features, the line profiles were extracted by applying the robust spatial profile of the H$\alpha$ and [OIII]$\lambda 5007$ features, respectively.  Flux calibration was then applied using a sensitivity function of the standard stars EG 131 and LTT 3864 with the IRAF tasks {\tt setairmass}, {\tt standard}, and {\tt sensfunc} and determining the response of the pixels corresponding to the emission line wavelengths.  Finally, the flux-calibrated line profiles were corrected for a Galactic foreground extinction of $E(B-V) = 0.116$ (Schlegel et al.\ 1998, Chornock et al.\ 2010).  Using the ratio of the H$\alpha$ and H$\beta$ features, we determined that host extinction was negligible across both of our host profiles and required no additional correction of the line profiles; this is in agreement with the results of Chornock et al.\ (2010), who also measured a negligible level of extinction in the explosion site HII region. Our final line profiles are plotted in Figure 1.

\section{Spatially-Resolved ISM Properties}
We used our extracted profiles of the H$\beta$, [OIII] $\lambda$5007, H$\alpha$, and [NII] $\lambda$6584 emission lines to construct emission line diagnostic ratio profiles for both the ``site" and ``host" observations. Our two diagnostic ratios of interest are [NII] $\lambda$6584/H$\alpha$ and [OIII] $\lambda$5007/H$\beta$. The [NII] $\lambda$6584/H$\alpha$ ratio is strongly correlated with metallicity (Veilleux \& Osterbrock 1987, Kewley et al.\ 2001), due to the dependence of the [NII] flux on primary and secondary nitrogen production (e.g. Chiappini et al.\ 2005, Mallery et al.\ 2007). However, Kewley \& Dopita (2002) also note that, due to the low ionization potential of [NII] $\lambda$6584, this ratio is also somewhat sensitive to the ionization parameter of the host environment\footnote{Here we define ionization parameter as the maximum velocity possible for an ionization front being driven by the local radiation field.}. Similarly, [OIII] $\lambda$5007/H$\beta$ is sensitive to the hardness of the ionizing radiation field, making it a useful tracer of the ionization parameter (Baldwin et al.\ 1981). While this ratio can also be double-valued with metallicity, it is far more sensitive to the ionization parameter at lower metallicities (Kewley et al.\ 2004). Since the ionization parameter itself is inversely dependent on metallicity (Dopita et al.\ 2006), these two ratios ultimately allow us to isolate metallicity as the fundamental physical parameter driving the evolution of our spatial profile. Another advantage to both of these diagnostic ratios is their insensitivity to reddening corrections, due to the close proximity of the emission lines being compared in each case (we found that effects from both the host and the foreground Galactic extinction were negligible in our analyses). With profiles for both the [NII] $\lambda$6584/H$\alpha$ and [OIII] $\lambda$5007/H$\beta$ diagnostic ratios, we used the Pettini \& Pagel (2004; hereafter PP04) calibration of the $O3N2$ metallicity diagnostic to construct metallicity profiles for the regions of the GRB 100316D host complex covered by our ``site" and ``host" observations. Finally, we use the H$\alpha$ fluxes in our data to calculate SFR profiles for our observations based on the relation of Kennicutt (1998).

The results of these analyses are plotted in Figure 1 (bottom). For our calculation of the emission line diagnostic ratios, PP04 metallicity, and SFR, we restricted our analyses to regions where all emission features had a $>5\sigma$ detection; this was generally limited by the flux of the [NII] $\lambda$6584 line. The data were also smoothed with a boxcar average of 5 pixels in the spatial direction, corresponding to the seeing conditions during the observations. It can be seen from these observations that GRB 100316D occurred in a region of the host complex with low metallicity and high SFR. From our spatial profiles and the position of the GRB-SN, we find a PP04 metallicity of log(O/H) + 12 = 8.2 $\pm$ 0.1 at the explosion site. This in excellent agreement with the metallicity determined in Chornock et al.\ (2011) based on an LDSS3 spectrum taken 3.3 days after the explosion at the GRB-SN explosion site. From examining both the ``site" and ``host" metallicity profiles, the host galaxy has a relatively flat metallicity profile, with a variation of approximately 0.3 dex and an average PP04 metallicity of log(O/H) + 12 $\approx$ 8.3 $\pm$ 0.1; these results also agree with the metallicity determined by Starling et al.\ (2011) in another region of the host complex. For both profiles, the minimum PP04 metallicity corresponds to the bright region of the host galaxy associated with the GRB 100316D/SN 2010bh explosion site. We also find a SFR of $\sim$1.7 M$_{\odot}$ yr$^{-1}$ at the host site, the highest SFR in the host complex. This is consistent with SFRs for other LGRB host galaxies and explosion sites (Levesque et al.\ 2010a, 2010b), although it should be noted that this is measured over a small $\sim 1" \times 1"$ region (corresponding to $\sim$1.3 kpc$^2$), whereas in other host studies the SFR is integrated over entire galaxies, with a much larger area ($\sim$10-30 kpc$^2$).

It should be noted that different strong-line metallicity diagnostics can produce drastically different metallicities from the same spectra and emission line fluxes, with differences ranging up to $\sim$0.8 dex (Kewley \& Ellison 2008). As a result, we restrict our analyses and comparisons to the PP04 metallicity diagnostic. For comparisons with metallicities determined from other diagnostics, the conversions of Kewley \& Ellison (2008) should be used.

\section{Discussion}
We have examined the host complex of GRB 100316D/SN 2010bh, constructing spatially-resolved emission line and metallicity profiles from spectrophotometric observations centered on both the explosion site and the brightest region of the host. Based on our results, we find little variation in metallicity across the host galaxy; however, we note that the GRB-SN explosion site is localized near a bright region of the host that represents a metallicity minimum and SFR maximum in our observations. LGRBs have long been associated with star-forming regions due to the young ages of their progenitors ($<$10Myr; Woosley et al.\ 2002). However, the lack of substantial variations in the host metallicity and the placement of the explosion site in the lowest-metallicity region of the host both have important implications for future studies of LGRB host galaxies.

In Figure 3 we compare host galaxy metallicities and explosion site metallicities for a small number of GRB host galaxies with spatially-resolved abundances. Using integral field unit spectroscopy of the dwarf SBc host galaxy of GRB 980425/SN 1998bw, Christensen et al.\ (2008) determined PP04 metallicities at 23 different sites across the host galaxy. Similar to the GRB 100316D/SN 2010bh host, they find a low average metallicity and no evidence of any statistically significant abundance gradient. They also note that the SN explosion site has a lower-than-average metallicity (indeed, only the neighboring Wolf-Rayet-rich region has a lower metallicity in their sample), although their measured abundance agrees with the overall host metallicity to within the errors. In another recent study, Th\"{o}ne et al.\ (2008) examined spatially-resolved ISM properties across the $z = 0.0889$ Sbc spiral host galaxy of GRB 060505, finding that the GRB site had a significantly lower metallicity than the other regions of the host. However, it should be noted that the classification of this particular GRB is a matter of hot debate; with a relatively short duration ($T_{90} = 4$ s) and no accompanying supernova detection despite its proximity, the nature of the GRB 060505 and its relation to other LGRBs remains unclear (Fynbo et al.\ 2006). Finally, Levesque et al.\ (2010c) determined PP04 metallicities for both the nucleus and explosion site within the GRB 020819 host galaxy. From Figure 3, we can see that for all four of these host galaxies, the explosion site metallicities show a good general agreement with the galaxies' average (or nucleus, in the case of GRB 020819) metallicities, displaying a small but uniform offset of $-0.1$ dex. These results suggest that global metallicities can indeed be treated as representative proxies for LGRB host environment metallicities in more distant hosts, where spatially-resolved studies are not always possible. Interestingly, this also sets the nearby LGRB host population apart from the hosts of Type Ibc supernovae, where large metallicity variations have been seen between the explosion sites and host nuclei (Modjaz et al.\ 2011). In addition, we find that for both the GRB 020819 and GRB 100316D/SN 2010bh host galaxies, our highest measured SFR is at the explosion site. Similarly, Christensen et al.\ (2008) find a higher luminosity-weighted SFR at the GRB 980425/SN 1998bw explosion site when compared to the galaxy as a whole or the mean luminosity-weighted SFR across the host (though other individuals sites do show higher SFRs), and Th\"{o}ne et al.\ (2008) find the highest luminosity-weighted SFR in the GRB 060505 host galaxy at the explosion site.

It is important to recognize that both GRB 980425/SN 1999bw and GRB 100316D/SN 2010bh represent an unusual class of LGRBs; these events are ``sub-luminous", with E$_{\gamma,iso}$ values that are much lower that the general LGRB population (such events are not even detectable at $z \gtrsim 0.3$), and show evidence of burst geometries that are quasi-spherical, in contrast to their tightly-beamed and more luminous counterparts (e.g. Soderberg et al.\ 2004, Soderberg 2006, Levesque et al.\ 2010e). This suggests that these events could be phenomenologically distinct from the general LGRB population. However, continued studies of these sub-luminous GRB-SNe and the ISM properties of their host galaxies could offer important insights into the nature of these events' progenitors and their connection with the larger sample of LGRBs.

Extending these spatially-resolved ISM studies to a wider range of nearby LGRB host galaxies would be extremely valuable. The host galaxies of GRB 020903 ($z = 0.251$, Ricker et al.\ 2002), GRB 030329/SN 2003dh ($z = 0.168$, Greiner et al.\ 2003), and GRB 060218/SN 2006aj ($z = 0.034$, Mirabal \& Halpern 2006) are all excellent candidates for these studies (see also Figure 8 of Starling et al.\ 2011). In addition, while GRB 020903 and GRB 060218 appear to belong to the same class of subluminous GRB/SNe as GRB 100316D/SN 2010bh (a supernova association for GRB 020903 was reported in Soderberg et al.\ 2005), GRB 030329/SN 2003dh has a notably higher luminosity and is often considered to be representative of the higher-redshift ``cosmological" class of LGRBs (e.g., Stanek et al.\ 2006); a spatially-resolved spectroscopic ISM study of this host galaxy could therefore prove particularly illuminating. Detailed maps of metallicity variations in all of these hosts would provide important data on the explosion sites of LGRBs, and further clarify how the metallicities of these sites compare to overall host metallicities. Additional properties such as ionization parameter, star formation rate, and young stellar population age would be useful for characterizing the typical explosion environments of LGRBs. Finally, stellar population synthesis studies at these sites and across the host galaxies would provide a unique opportunity to compare the different stellar populations that dominate the host light and give rise to the progenitors of these rare and intriguing events.

We are grateful for the hospitality and assistance of the support staff at Las Campanas Observatory in Chile. This paper utilized data from the Gamma-Ray Burst Coordinates Network (GCN) circulars. EML is supported by NASA through Einstein Postdoctoral Fellowship grant number PF0-110075 awarded by the Chandra X-ray Center, which is operated by the Smithsonian Astrophysical Observatory for NASA under contract NAS8-03060. GRB research at Harvard is supported in part by Swift AO5 grant 5080010 and AO6 grant 6090612.

\clearpage
\begin{figure}
\epsscale{0.42}
\hspace{11pt}
\plotone{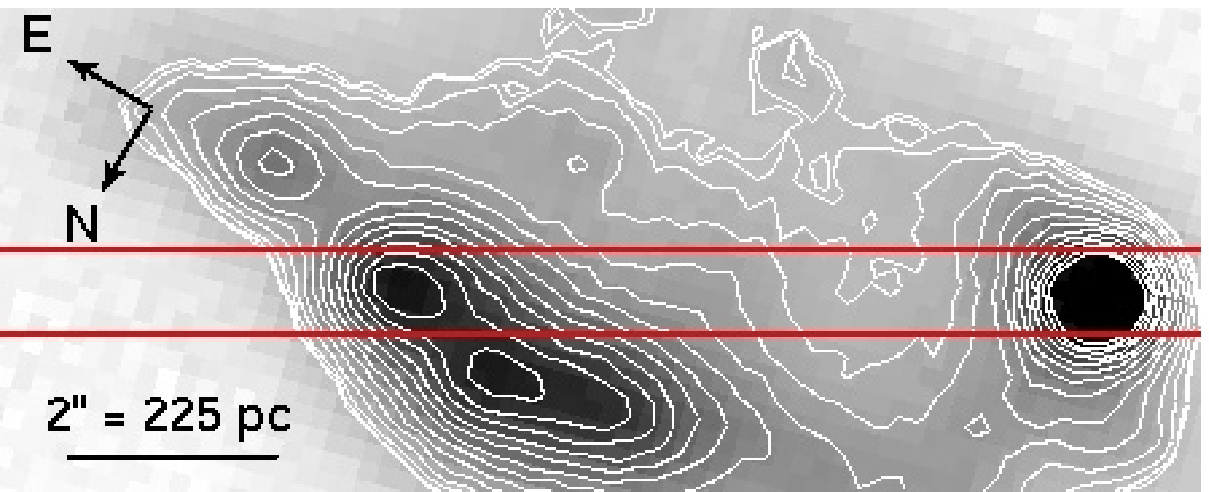}
\hspace{32pt}
\plotone{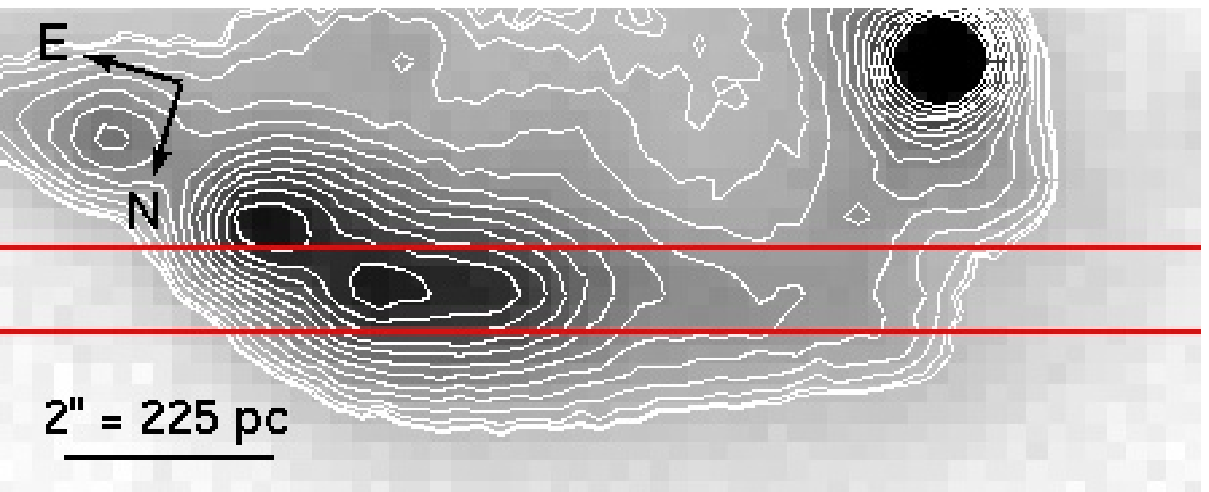}
\epsscale{0.49}
\plotone{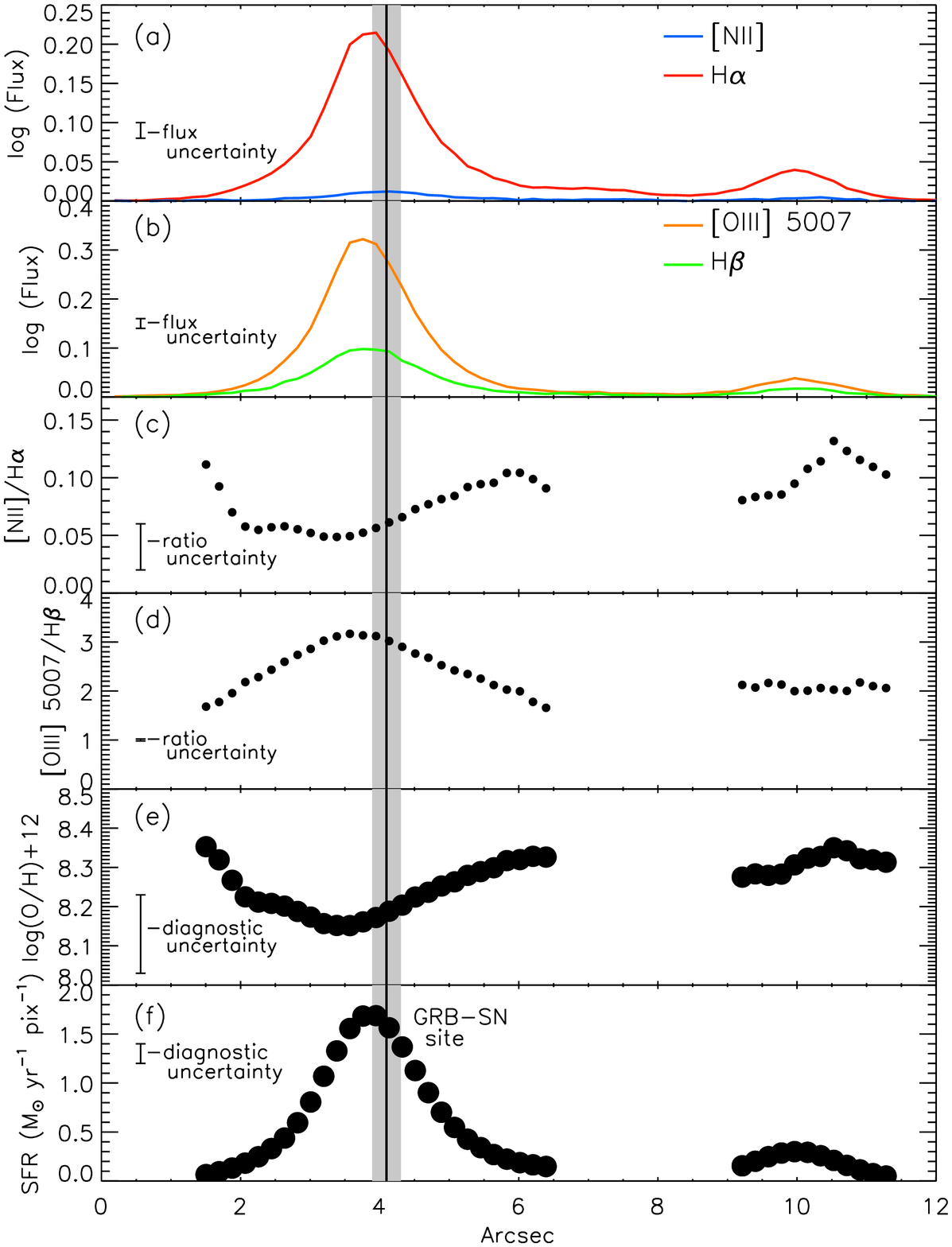}
\plotone{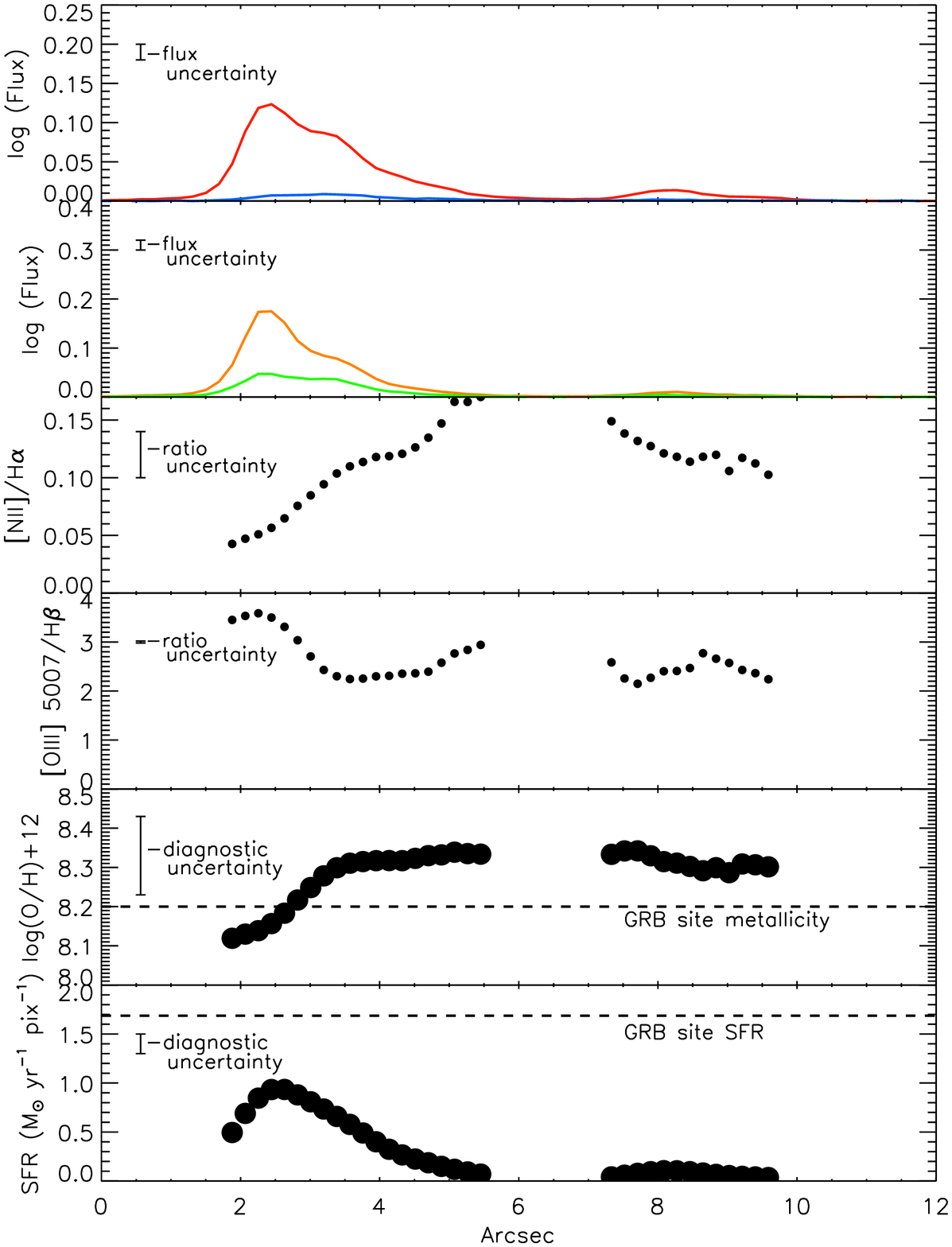}
\caption{Top: Acquisition images showing position of the 1" slit for our observations of the GRB 100316D/SN 2010bh explosion site (left) and host galaxy (right). The images were taken in the g-band, with contours indicating g-band brightness. Bottom: Emission line fluxes, ratios, and metallicity profiles for our spectra of the explosion site (left) and host galaxy (right). From top to bottom, these show (a) the H$\alpha$ and [NII] $\lambda$6584 line profiles, (b) the H$\beta$ and [OIII] $\lambda$5007 line profiles, (c) the [NII] $\lambda$6584/H$\alpha$ ratio, (d) the [OIII] $\lambda$5007/H$\beta$ ratio, (e) metallicity determined based on the {\it O3N2} diagnostic calibration of Pettini \& Pagel (2004), and (f) SFR determined based on the H$\alpha$ diagnostic of Kennicutt (1998). On the left, the GRB-SN position is indicated by the solid black vertical line, with errors illustrated by the gray region. On the right, the observed metallicity (log(O/H) + 12 $\sim$ 8.2) and SFR ($\sim$1.6 M$_{\odot}$ yr$^{-1}$) of the GRB-SN is given by the dashed horizontal line in panels (e) and (f), respectively. In panels (a) and (b) flux is in units of 10$^{-13}$ ergs cm$^{-2}$ s$^{-1}$.} 
\end{figure}
\clearpage

\begin{figure}
\epsscale{0.9}
\plotone{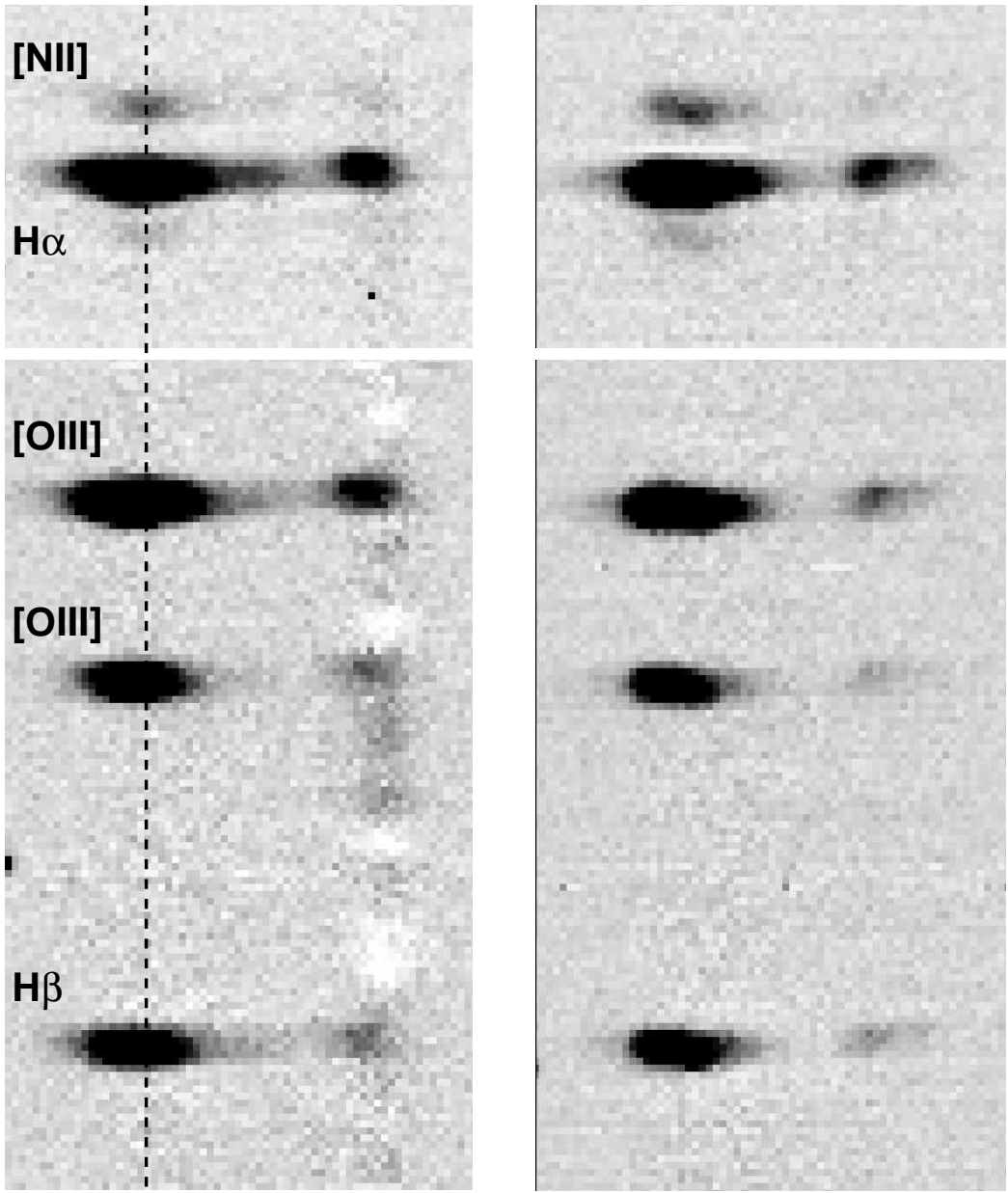}
\caption{Background-subtracted 2D spectra of our ``site" (left) and ``host" (right) spectra, showing the [NII] $\lambda$6584 and H$\alpha$ features (top) and the [OIII] $\lambda$5007,4959 and H$\beta$ features (bottom). In the ``site" spectrum, the explosion site of GRB 100316D/SN 2010bh is indicated by the dashed black line. The negative residuals in the ``site" 2D spectrum are due to subtraction of the bright foreground star (seen in Figure 1), and have no impact on the derived line intensities.}
\end{figure}

\begin{figure}
\epsscale{1}
\plotone{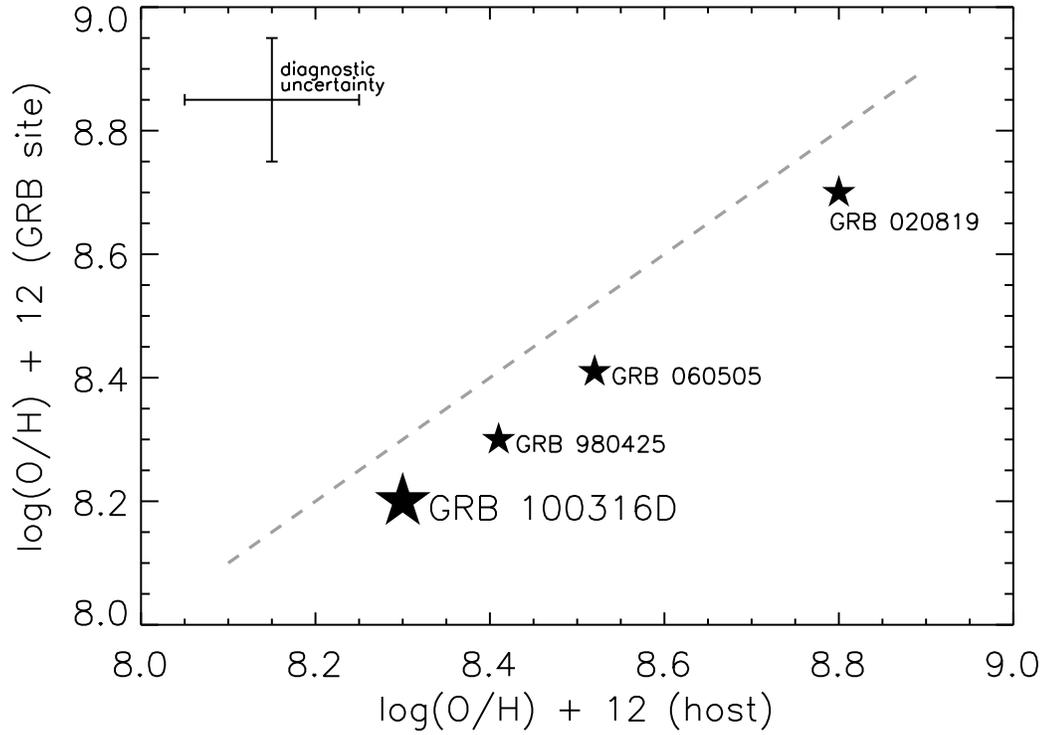}
\caption{Explosion site metallicities vs. average host metallicities for a sample of four nearby GRB host galaxies. Metallicities for GRB 980425 and GRB 060505 are taken from Christensen et al.\ (2008) and Th\"{o}ne et al.\ (2008), respectively, and were determined using the Pettini \& Pagel (2004) {\it O3N2} diagnostic calibration. The ``host" metallicities here are taken as averages of metallicity measurements made throughout the host galaxies. Metallicities for GRB 020819 was taken from Levesque et al.\ (2010c) and determined using the Pettini \& Pagel (2004) {\it N2} diagnostic calibration that is better suited for higher-metallicity environments. The ``host" metallicity in this case is the metallicity of the host galaxy nucleus. Metallicities for GRB 100316D are taken from this work. All four host environments fall $\sim$0.1 dexbelow the theoretical relation where explosion site metallicity and host metallicity are identical, plotted here as a gray dashed line, though this is within the uncertainty of the Pettini \& Pagel (2004) diagnostics.}
\end{figure}

\end{document}